\documentclass[10pt]{wlscirep} 
\usepackage[utf8]{inputenc}
\usepackage[T1]{fontenc}
\usepackage{braket}
\usepackage{subcaption}
\usepackage{float}
\usepackage{dblfloatfix}
\usepackage{placeins}
\usepackage{adjustbox}
\usepackage{tabularx}
\usepackage{seqsplit}
\usepackage{multirow}
\usepackage{array}
\usepackage{soul}

\usepackage{colortbl}
\usepackage{pgfplots}
\usepackage{pgfplotstable}
\usepackage{pgf} 
\usepackage{etoolbox}
\definecolor{high}{HTML}{08ef15}  
\definecolor{low}{HTML}{2f5652}  
\newcommand*{\opacity}{50}
\newcommand*{\minval}{0.0}
\newcommand*{\maxval}{2.5}
\newcommand{\gradient}[1]{
    \small
    \ifdimcomp{#1pt}{>}{\maxval pt}{#1}{
        \ifdimcomp{#1pt}{<}{\minval pt}{#1}{
            \pgfmathparse{int(round(100*(#1/(\maxval-\minval))-(\minval*(100/(\maxval-\minval)))))}
            \xdef\tempa{\pgfmathresult}
            \cellcolor{high!\tempa!low!\opacity} #1
    }}
}

\usepackage{bm}
\makeatletter
\AtBeginDocument{\DeclareMathVersion{bold}
\SetSymbolFont{operators}{bold}{T1}{times}{b}{n}
\SetMathAlphabet{\mathrm}{bold}{T1}{times}{b}{n}
\SetMathAlphabet{\mathit}{bold}{T1}{times}{b}{it}
\SetMathAlphabet{\mathbf}{bold}{T1}{times}{b}{n}
\SetMathAlphabet{\mathtt}{bold}{OT1}{pcr}{b}{n}
\SetSymbolFont{symbols}{bold}{OMS}{cmsy}{b}{n}
\renewcommand\boldmath{\@nomath\boldmath\mathversion{bold}}}
\makeatother

\def\BibTeX{{\rm B\kern-.05em{\sc i\kern-.025em b}\kern-.08em
    T\kern-.1667em\lower.7ex\hbox{E}\kern-.125emX}}
\pgfplotsset{compat=1.18}

\title{Exploring Quantum Control Landscape and Solution Space Complexity through Optimization Algorithms \& Dimensionality Reduction}

\author[1,3,*]{Haftu W. Fentaw}
\author[2,3]{Steve Campbell}
\author[1,3]{Simon Caton}
\affil[1]{School of Computer Science, University College Dublin, Dublin, Ireland}
\affil[2]{School of Physics, University College Dublin, Dublin, Ireland}
\affil[3]{Centre for Quantum Engineering, Science, and Technology, University College Dublin, Dublin, Ireland}
\affil[*]{haftu.fentaw@ucdconnect.ie}

\keywords{
Genetic Algorithms (GA), Principal Component Analysis (PCA), Quantum Control Landscape (QL), Reinforcement Learning (RL), Stochastic Gradient Descent (SGD)}

\begin{abstract}
Understanding the quantum control landscape (QCL) is important for designing effective quantum control strategies. In this study, we analyze the QCL for a single two-level quantum system (qubit) using various control strategies. We employ Principal Component Analysis (PCA), to visualize and analyze the QCL for higher dimensional control parameters. Our results indicate that dimensionality reduction techniques such as PCA, can play an important role in understanding the complex nature of quantum control in higher dimensions. Evaluations of traditional control techniques and machine learning algorithms reveal that Genetic Algorithms (GA) outperform Stochastic Gradient Descent (SGD), while Q-learning (QL) shows great promise compared to Deep Q-Networks (DQN) and Proximal Policy Optimization (PPO). Additionally, our experiments highlight the importance of reward function design in DQN and PPO demonstrating that using immediate reward results in improved performance rather than delayed rewards for systems with short time steps. A study of solution space complexity was conducted by using Cluster Density Index (CDI) as a key metric for analyzing the density of optimal solutions in the landscape. The CDI reflects cluster quality and helps determine whether a given algorithm generates regions of high fidelity or not. Our results provide insights into effective quantum control strategies, emphasizing the significance of parameter selection and algorithm optimization.
\end{abstract}
\begin{document}

\flushbottom
\maketitle

\thispagestyle{empty}

\section{Introduction}

\label{sec:introduction}
As the demand for advanced computational capabilities continues to rise, there is a growing need for alternatives to solve complex problems that are either beyond the reach of classical computers or could use a boost with improved computational capability. Quantum computing has emerged as a viable alternative with the potential to tackle certain classes of problems currently not amenable with today's classical algorithms, such as factoring large numbers (Shor's algorithm) \cite{Shor_1997}, or speeding up certain algorithms as in quantum machine learning \cite{Biamonte_2017}. However, the inherent fragility of quantum states means that to harness this potential, effective quantum control strategies are essential for manipulating quantum states. In order to design an effective control strategy it is therefore important to understand the quantum control landscape (QCL). 

The QCL refers to the multidimensional space of control variables and represents the relationship between the control parameters of a quantum system and the associated performance measures, such as the distribution of candidate optimal solutions in any local neighborhood or the trajectories through control space to the optimal solution~\cite{C4CP03853C, 9502048}. The study of QCLs is critical for understanding the controllability, and corresponding optimality, that is achievable by manipulating the quantum systems using, e.g., external control fields. Exploration of this landscape helps in identifying the optimal path a quantum system takes during a transition to a target state \cite{Chakrabarti_2007}. By varying the properties of the control pulses (i.e. applying variable external fields), the state of the system can be driven to a desired state from its initial configuration~\cite{PhysRevA-84-012109}. 

For a closed system, the process during which this change in the state occurs is governed by the time-dependent Schr\"odinger equation
\begin{equation}
    \label{eq:SE}
    i\hbar\frac{\partial}{\partial t}\ket{\psi(t)} = H(t)\ket{\psi(t)}
\end{equation}
where $H(t)$ is the Hamiltonian of the system which describes the total energy and $\ket{\psi(t)}$ is the time evolved state of the system. In general, solving~\eqref{eq:SE} is a difficult problem for an arbitrary time-dependence. One approach is to determine $\ket{\psi(t)}$ using a discrete time analysis by assuming the overall evolution time, $T$, is divided into $N$ equal partitions of size $\Delta t=T/N$ such that the state at time $t+\Delta t$ is given by
\begin{equation}
\ket{\psi(t+\Delta t)} = e^{-\frac{i}{\hbar}H(t)\Delta t} \ket{\psi(t)}
\label{timeEvolution}
\end{equation}

In quantum control problems it is common to express the time-dependence in the Hamiltonian, $H(t)$ as
\begin{equation}
    H(t) = H_d + u(t)H_c  
    \label{Hamiltonian}
\end{equation}
where $H_d$ is the drift Hamiltonian and $u(t) H_c$ is the control Hamiltonian. The objective of quantum control is, therefore, to find a $u(t)$ that will maximize the probability of achieving some target objective, e.g. evolving to a given final state~\cite{Larocca_2018, Koch2022-bo}. The cost function for the control is then given by the quantum state fidelity
\begin{equation}
    F = |\braket{\psi(T)|\Psi}|^2
    \label{fidelity}
\end{equation}
where $\ket{\psi(T)}$ is the state at the end of the evolution and $\ket{\Psi}$ is the desired target state.

The QCL allows us to analyze the complexity of the solution space by plotting the combination of discrete field values that will result in high target state fidelity values. While for simple piece-wise constant pulses consisting of only two segments, i.e. $N=2$, 
 the full QCL can be plotted, in order to visually inspect the complexity of the solution space for higher dimensions, i.e. $N>2$, we need to resort to the use of \textit{dimensionality reduction} techniques.

Finding the right combination of control parameters which will result in achieving a desired target is a complex task that requires the use of different optimization algorithms. There is a rich body of work exploring different approaches, including notable examples such as Gradient Ascent Pulse Engineering (GRAPE)~\cite{KHANEJA2005296}, stochastic gradient descent (SGD)-based works such as~\cite{Ferrie190404}, and reinforcement learning (RL)-based approaches as demonstrated in \cite{Zhang_2019}.  

\begin{figure}[htbp]
    \centering
        \includegraphics[width=0.9\textwidth]{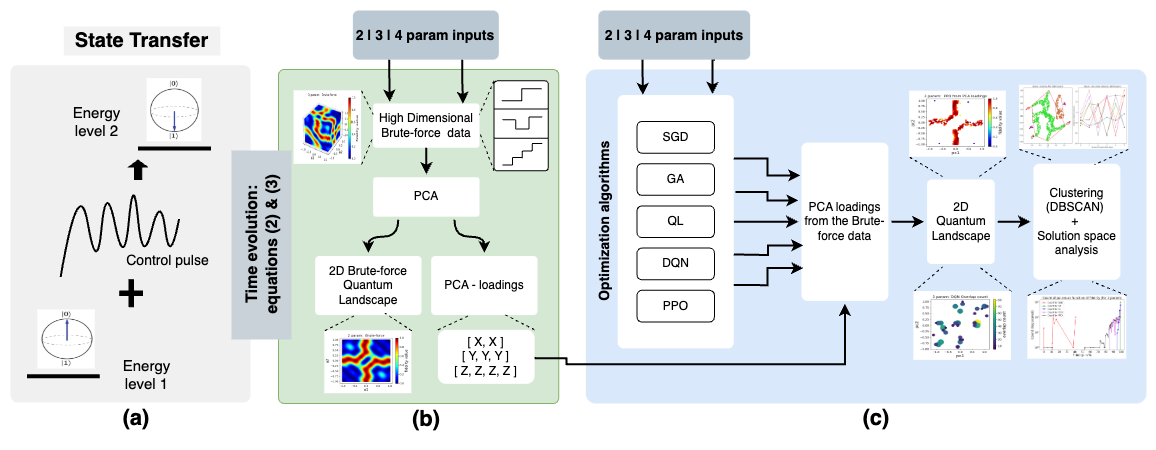} 
    \caption{High level diagram: The basic steps involved in this work: (a) the state transfer problem, (b) the brute-force solution (by converting the continuous control signal into discrete values ), (c) optimization algorithms based solution}
    \label{fig:highleveldiagram}
\end{figure}

In this work, the performance and suitability of traditional optimization algorithms, such as SGD, genetic algorithms (GA), and RL algorithms, for exploring the QCL are presented. The high level description of the state transfer problem and steps followed in our work is presented in Fig.~\ref{fig:highleveldiagram}. We focus on the control of a single two-level quantum system (qubit) initially in state $\ket{0}$ and aim to perform a bit flip operation to state $\ket{1}$ as seen in Fig.~\ref{fig:highleveldiagram}(a) using a control pulse. The qubit's evolution is governed by~\eqref{timeEvolution} and \eqref{Hamiltonian}, leading to a piece-wise constant control signal, as illustrated in Fig.~\ref{fig:highleveldiagram}(b). The QCL for different input combinations is generated using a brute-force approach (for later comparison), the results for each parameter are passed through a 2 component PCA to obtain a 2D landscape and the PCA loadings (which will be used in later stages of the analysis). Details of this are presented in section~\ref{sec:dimensionalityReduction}. Following this, Fig.~\ref{fig:highleveldiagram}(c), shows the experiments we perform using different optimization algorithms to generate the control pulses (we refer to section~\ref{sec:experimentalsetup}). After the control pulses are generated,  they are passed through PCA (using PCA loadings from previous step) to obtain the 2D landscape. Once the 2D landscapes are plotted, we perform a number of analyses to better understand the complexity of the landscape, and find the algorithms which can explore the landscape well. Details of this can be found in sections~\ref{sec:results} and~\ref{sec:solutionspacecomplexity}.

The main contributions of this work are:
\begin{itemize}
    \item Investigation of how well different algorithms explore the solution space / landscape with the view to characterize the potential suitability of different techniques. 
    \item Introduced Dimensionality Reduction (DR) techniques to visualize and study higher dimensional quantum control landscapes
    \item Designed proper RL networks and reward signals specifically for short time step (small steps per episode) quantum control problems
    \item Analysis of the solution space complexity - to better understand how easy or difficult it is to find high fidelity solutions in certain regions of the landscape and which algorithms are best suited for this. 
\end{itemize}

\section{Prior Related work}
\label{sec:relatedwork}
The study of QCLs is well established at this point and one of the key insights we can extract from QCL analysis is whether the landscape is trap-free or it contains local minima that can hinder optimization~\cite{PhysRevA-84-012109}. It therefore follows that understanding the structure of the landscape can help avoid these traps and improve the efficiency of control strategies~\cite{PhysRevA-84-012109}. When no constraints are imposed on the available controls for the system, the QCL is devoid of local minima and is trap-free~\cite{Russell_2017}. However, in many real-world scenarios, some constraints are unavoidable. As shown, for example, in~\cite{Larocca_2020_PhysRevA.101.023410}, the QCL is not trap-free in most practical situations where constraints are imposed on the system, making it necessary to use more sophisticated optimization techniques to escape traps. 

In Ref.~\cite{Larocca_2018}, the authors conducted a detailed analysis of the QCL for a single qubit bit-flip using a two-parameter control system for various evolution times, both below and above the so-called quantum speed limit~\cite{Deffner2017}. The quantum speed limit, $T_{min}$ sets a fundamental lower bound on the time necessary to connect two quantum states and for a single qubit its value can be analytically determined as in~\cite{Hegerfeldt_2013, Poggi2013}. Their findings indicate that for $T/T_{min}<1$, the landscape is relatively simple, featuring only a single global maximum but importantly with a fidelity, $F<1$. As the total evolution time is increased well beyond the quantum speed limit, the landscape becomes more complex and there are a number of local maxima which are able to achieve perfect state transfer. We extend their work to explore details of the landscape in higher dimensions, and how the possible ridges and valleys in higher dimensions can influence the solution space.

It is worth highlighting that the QCL can also reveal important physical insights into the underlying quantum dynamics. For example, the presence of broad plateaus or ridges in the landscape may suggest that a wide range of control parameters can achieve near-optimal performance, highlighting the robustness of certain quantum systems to noise and perturbations\cite{PhysRevA.86.052117, PhysRevA-84-012109}. On the contrary, if a quantum system has steep landscapes, this suggests that the system is highly sensitive and it requires more precise control strategies to find the optimal control pulses. 

In order to study the details of the QCL at higher dimensions, a technique to reduce the number of dimensions or features of a dataset while retaining as much information as possible is necessary. Dimensionality reduction algorithms such as Principal Component Analysis (PCA)~\cite{MACKIEWICZ1993303}, t-distributed Stochastic Neighbor Embedding (t-SNE)~\cite{JMLR:v9:vandermaaten08a} and Uniform Manifold Approximation and Projection for Dimension Reduction~(UMAP)~\cite{mcinnes2020umapuniformmanifoldapproximation} are among the most commonly used techniques. The decision of choosing which algorithm to use depends on several factors including how sensitive the algorithm is to small changes in its parameters and how well the algorithm preserves the structure of the data after applying it. There are studies into these details (e.g. \cite{Huang2022_comprehensive_DR}) that can assist in this decision process.  For example, authors in Ref.~\cite{berger2024dimensionalityreductionclosedloopquantum} applied dimensionality reduction in quantum gate calibration, PCA is used in Ref.~\cite{li2018visualizinglosslandscapeneural} to visualize the loss landscape in neural networks, and in Ref.~\cite{fan2024manifoldconnectednessquantumcontrol} PCA is used to project quantum control trajectories into three dimensions. 
Both t-SNE and UMAP require tuning parameters that significantly influence the results, often leading to inconsistent outcomes that can be difficult to interpret~\cite{10.5555/3546258.3546459}. For instance, t-SNE relies on parameters such as perplexity, learning rate, and the number of iterations, while UMAP depends on the number of neighbors and minimum distance between points. In contrast, PCA requires only the number of principal components to be specified, making it simpler to use and resulting in more stable and interpretable outcomes. For these reasons (and also based on initial experimentation with each of these methods), we will focus on PCA to visualise the QCL. 

The QCL is typically studied using various optimization techniques to find the best control parameters that guide the system. Traditional control algorithms including Gradient-based methods, Genetic Algorithms (GA) and Bang-Bang Control are widely used for navigating the QCL and optimizing control parameters. While the authors in Ref.~\cite{Ferrie190404} introduced a central difference based gradient approach (self guided quantum tomography) and argued this iterative approach is more efficient than other methods, those in Ref.~\cite{Brown_2023} utilized genetic algorithms for optimal quantum state control and claimed they observed fast preparation times, and resilience to noise when using genetic algorithms. The drawbacks of gradient-based algorithms is that they can get trapped in local minima, especially in systems with complex landscapes, and their sensitivity to initial conditions~\cite{Netrapalli2019-gu}. While GAs can avoid local minima more effectively than gradient-based methods, they may still suffer from inefficiencies in high-dimensional search spaces~\cite{10.1371/journal.pone.0303088}. Authors in Ref.~\cite{Barnes2015-zj} introduced an analytical robust quantum control approach that not only yields explicit constraints on the control field but also ensure that the leading-order noise-induced errors in a qubit’s evolution cancel exactly.

In recent years, Reinforcement Learning (RL) based approaches have emerged as alternatives for quantum optimal control. The major advantage of RL is its ability to explore vast control landscapes autonomously, finding creative control solutions that may be difficult for traditional methods to uncover. For instance, authors in Ref.~\cite{s41534-019-0141-3} used trusted-region-policy-optimization (TRPO) RL algorithms to train agents for the control of two-qubit unitary gates by adding control noise into training environments and they found the agent demonstrates a two-order-of-magnitude reduction in average-gate-error compared to gradient based methods. This shows that RL algorithms are robust to noise and model uncertainties because they can adapt to noisy environments by learning policies that are more resilient to fluctuations in control parameters, making them practical for real-world quantum systems where noise and environmental factors affect control fidelity. RL algorithms can also handle high-dimensional control spaces and complex quantum dynamics by learning from interactions with the environment, thus reducing the need for precise system modeling. In Ref.~\cite{PhysRevX.12.011059}, the authors trained a model-free RL agent which learns through trial-and-error interaction with the quantum system and hence there is no need for the precise modeling of the physical system. They concluded that this is of immediate relevance allowing the adaptation of quantum control policies to the specific system in which they are deployed and complete elimination of model bias. 

Different RL based methods were compared with traditional optimization techniques in Ref.~\cite{Zhang_2019}, where the authors demonstrated that RL algorithms (TQL~\cite{sutton1999reinforcement}, DQN~\cite{mnih2015human} and PG~\cite{NIPS1999_464d828b}) can outperform traditional methods (SGD~\cite{Ferrie190404} and Krotov~\cite{krotov1996global}) when the problem size is scaled up. The authors showed that for a relatively small number of iterations (up to 500) SGD would fail to reach optimal solutions compared with ML techniques, however it is worth noting that allowing SGD more iterations can result in improved performance as detailed in section \ref{sec:experimentalsetup}.

Despite RL-based algorithms showing better performance, there are drawbacks to using RL in quantum control. RL methods often require a large number of training episodes, particularly for complex quantum systems. Additionally, RL systems are sensitive to hyper-parameters, such as the learning rate and exploration-exploitation trade-off, which must be carefully tuned for optimal performance~\cite{sutton1999reinforcement}. In this work, an effort is made to answer the question of when one should consider using RL techniques, which RL methods are beneficial, and which are excessive by visualizing the landscape and studying the solution space complexity.

Indeed, the fact that there are different options for quantum control notwithstanding, there has been limited research into visualizing and understanding the QCL in higher-dimensional parameter spaces. Therefore, this work contributes by providing: (a) a framework for the exploration and visualization of QCL in such complex spaces and (b) algorithms that we can utilize to get optimal control pulses, offering new insights and tools for optimizing quantum control.

\section{Visualizing Higher Dimensional Landscapes Using Principal Component Analysis (PCA)}
\label{sec:dimensionalityReduction}
PCA is a statistical technique for dimensionality reduction that transforms a higher dimensional data set into a lower dimensional representation. It seeks to maximise the amount of information retained while reducing the dimensionality of the data. By projecting data onto a new set of orthogonal axes, called principal components, PCA identifies the directions of maximum variance in the data. The first principal component accounts for the largest possible variance, with each subsequent component capturing the next highest variance under the constraint of being orthogonal to the preceding components. This process helps in extracting key features and making it easier to visualize and analyze high-dimensional datasets.


By applying PCA on 3-parameter or 4-parameter quantum landscapes, we can create 2-dimensional visualisations of the landscapes which can allow us to gain insight into the complexities and characteristics of the QCL for the Hamiltonian. To make things concrete, in this work we focus on the QCL generated for $N$-parameter pulses, i.e. the continuous control field $u(t)$ over total evolution time $T$ is divided into $N$ equal partitions (time steps), each partition with its corresponding amplitude, resulting in a piece-wise constant function for $u(t)$. 
\begin{figure*}[htbp]
    \centering
    \begin{subfigure}[b]{0.295\textwidth}
        \centering
        \includegraphics[width=\textwidth]{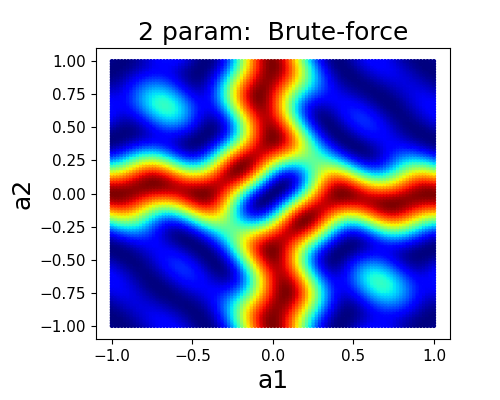} 
        \caption{}
        \label{fig:2paramraw}
    \end{subfigure}
    \begin{subfigure}[b]{0.295\textwidth}
        \centering
        \includegraphics[width=\textwidth]{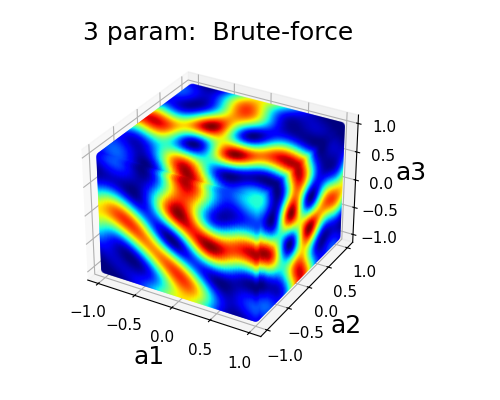} 
        \caption{}
        \label{fig:3paramraw}
    \end{subfigure}
    \begin{subfigure}[b]{0.295\textwidth}
        \centering
        \includegraphics[width=\textwidth]{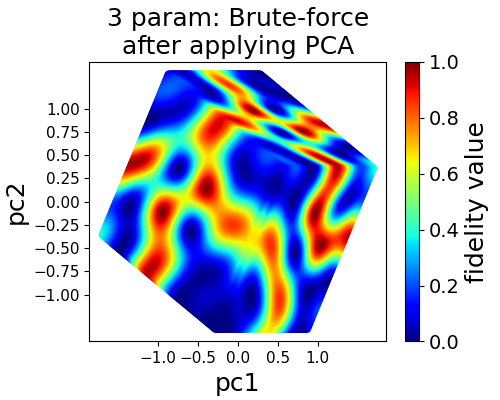} 
        \caption{}
        \label{fig:3paramPCAraw}
    \end{subfigure}
    \caption{Quantum landscape using brute-force data: (a) 2 parameter~\cite{Larocca_2018},  (b) 3 parameter, and (c) 3 parameter after applying PCA. Data points were created by generating every possible combination of values in the range [-1, 1] for each axis, with each axis divided into 100 intervals. Axis labels a1, a2 and a3 represent parameters 1, 2 and 3 in each axis, pc1 and pc2 represent the first two principal components.}
    \label{fig:bruteforce}
\end{figure*}
We consider the evolution of a single qubit with $T/T_{min}=2$, initial state of $\ket{0}$ and target state of $\ket{1}$ where $\ket{0}$ and $\ket{1}$ are the eigenstates of $\sigma_z$ and assume the system is governed by the Hamiltonian      
\begin{equation}
\label{eq:LZ}
    H(t) = \frac{\sigma_{x}}{2} + 2u(t)\sigma_{z}
\end{equation} 
The Hamiltonian in Eq.~\eqref{eq:LZ} is closely related to the celebrated Landau-Zener model which captures a remarkably diverse range of physical settings, as discussed, for example in Ref.~\cite{IVAKHNENKO20231} and we refer to Refs.~\cite{Larocca_2018, Hegerfeldt_2013} where similar control techniques as will be considered below were employed.

\begin{figure}[htbp]
    \centering
    \begin{subfigure}[b]{0.275\textwidth}
        \centering
        \includegraphics[width=\textwidth]{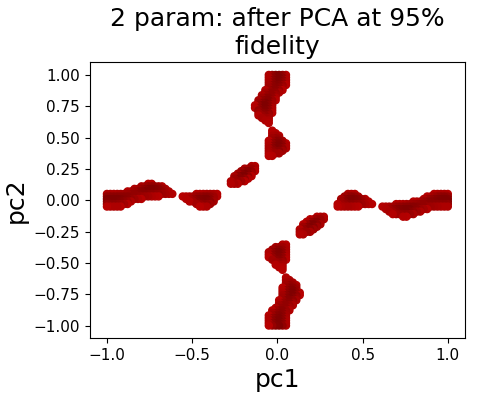} 
        \caption{}
        \label{fig:2paramPCAhighfid}
    \end{subfigure}
    \begin{subfigure}[b]{0.275\textwidth}
        \centering
        \includegraphics[width=\textwidth]{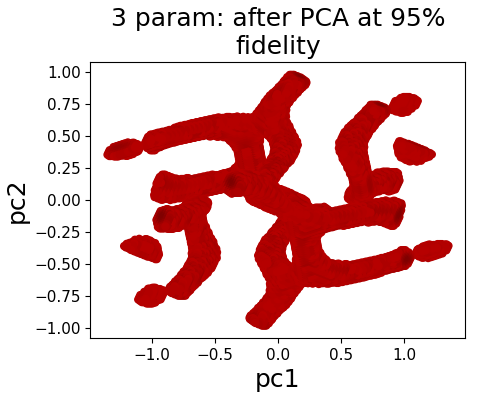} 
        \caption{}
        \label{fig:3paramPCAhighfid}
    \end{subfigure}
    \begin{subfigure}[b]{0.275\textwidth}
        \centering
        \includegraphics[width=\textwidth]{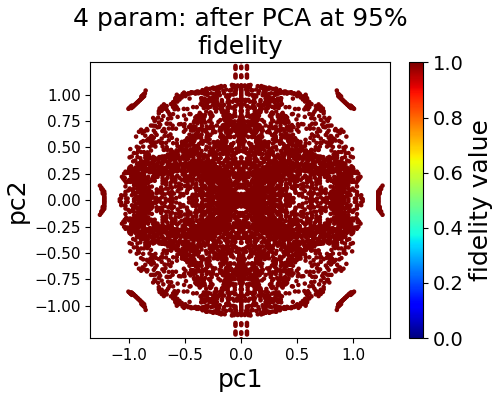} 
        \caption{}
        \label{fig:4paramPCAhighfid}
    \end{subfigure}
    \caption{Quantum control landscape after applying PCA and extracting regions of high fidelity (fidelity above 0.95): (a) Two parameter, (b) Three parameter, and (c) Four parameter, the dataset which is used to generate these plots is generated using a brute-force combination of inputs in the range [-1, 1]}
    \label{fig:afterapplyingPCAhighfid}
\end{figure}

The two parameter and three parameter quantum landscapes resulting from using brute-force combination of 100 values for each parameter in the range [-1, 1] are shown in Fig.~\ref{fig:2paramraw} and Fig.~\ref{fig:3paramraw}. As can be seen from Fig.~\ref{fig:3paramraw}, details of the landscape for the 3 parameter case are hard to see. (Note: throughout this paper, the axis labels a1, a2, and a3 represent parameter 1, parameter 2, and parameter 3, respectively, while pc1 and pc2 denote principal component 1 and principal component 2). In order to see underlying properties of the landscape for higher dimensions, we apply PCA with two principal components on the raw brute-force data points which results in the  landscapes shown in Fig.~\ref{fig:3paramPCAraw}. The landscapes in Fig.~\ref{fig:bruteforce} show both the high fidelity (red) and low fidelity (blue) regions. As the target is to achieve a high-fidelity state transfer, we can gain more insight into the space of effective control protocols by focusing on only regions with high target state fidelity, i.e. $F>0.95$, shown in Fig.~\ref{fig:afterapplyingPCAhighfid}.

 As it is evident from the PCA representation of the landscapes in Fig.~\ref{fig:afterapplyingPCAhighfid}, higher dimensional landscapes are filled with more high fidelity regions as compared to the low dimensional landscapes. From this observation, it is clear that by dividing the total evolution time $T$ in to more partitions (hence high dimensional landscape), we increase the chances of finding more control signals that will result in high fidelity (complete population transfer).  \par 
 
Clearly, we cannot simply try the brute-force combination of inputs for more than a small number of parameters and hope we can land in a combination which will result in high fidelity. Instead, in order to generate a control pulse with reasonably good fidelity, we use optimization algorithms such as SGD, GA, QL, DQN or PPO to find a control signal which will drive our system into its target state effectively.  Details of these algorithms and the experimental setups for each experiment are discussed next.

\section{Experimental Setup and Computational Techniques}
\label{sec:experimentalsetup}
We can study the QCL generated when using SGD, GA, QL, DQN and PPO for $N$-parameter optimization using~\eqref{eq:LZ}. In order to generate the QCLs as a function of fidelity, we run 1000 experiments for each algorithm, hence generated 1000 data-points, we then pass the results through a 2 component PCA to visualize the landscape in 2D. Since the results of the PCA will depend on the number of data-points available, we first generated PCA loadings for 2, 3, and 4 parameters using data points created by the brute force combination of possible values between [-1, 1]. PCA loadings represent the coefficients assigned to the original variables when forming the principal components and they indicate the contribution of each original feature to the principal components. These PCA loadings are then used to transform the 2, 3 and 4 parameter results from the above five algorithms to a 2-component PCA transformation. In our study, Numpy and Qutip~\cite{JOHANSSON20131234} libraries are used to represent the quantum states and other related vectors and matrices resulting during the optimization process. A minimum infidelity (1-fidelity) value  of 0.001 is used, and when the infidelity value from a certain algorithm is below this value, we assume the target is achieved and the algorithm exits the optimization loop immediately. A brief summary of the algorithms used in this work is presented below.

\subsection{SGD with Momentum}
Stochastic Gradient Descent (SGD) is an optimization algorithm that is used to minimize the cost function (difference between target value and calculated value) in optimization problems. In the context of quantum control, SGD can be used to generate control pulses that would drive the initial state towards the target state as much as possible. Starting from a list of random values whose length is equal to the number of parameters, $N$, these values are iteratively updated by generating a set of infinitesimally small values which are added to or subtracted from the initial random values, until a pulse that can result in high fidelity (above the target fidelity threshold) is found or the total number of iterations are exhausted. 

To improve the convergence time of the traditional central difference based SGD algorithm used in \cite{Ferrie190404} and \cite{Zhang_2019}, we can add a momentum term. This addition results in improved performance when compared with the implementation without momentum. A learning rate of 0.01, and a momentum of 0.95 are used. The initial pulses consist of a set of random values between -1 and 1, with a length equal to the number of parameters $N$. Unlike the 500 iterations used in \cite{Zhang_2019}, a maximum iteration of 10,000 is used as the stopping criteria for the SGD so that the algorithm has enough time to find optimal results and avoid sub-optimal regions. If the algorithm is unable to find an optimal control pulse after the max iterations are exhausted, the control pulse from the final iteration is returned. 

\subsection{Genetic Algorithm (GA)}
Genetic algorithms (GA), a class of optimization techniques inspired by the principles of natural selection and genetics, operate by evolving a population of candidate solutions through iterative processes of selection, crossover, and mutation, aiming to improve the solutions over successive generations based on fitness criteria. 

In the context of quantum control, the genes will be the possible pulse amplitudes (we limit number of genes to be 100 in the range to [-1, 1] in our experiment), the chromosome will represent the complete control pulse of length $N$, the population will be the collection of the chromosomes, and we use a mutation rate of 0.3. The length of each chromosome will be equal to the number of parameters we are interested in. The fitness function is, naturally, the fidelity of the system after the selected chromosome is tested for its performance using the given Hamiltonian. In every iteration, chromosomes that are among the top 30 percent of the population (based on their fitness value) and those that are in the last 20 percent are selected to be part of the next generation of the population. The decision to include those in the bottom 20 percent is to enable the algorithm have enough genetic variations in the cross over and mutation stages of the algorithm. To keep the total population constant, the remaining 50 percent of the population is accounted for by generating new population through crossover and mutation. The initial population and chromosomes are generated by selecting genes randomly.

In our experiments, we limit the algorithm to have a maximum of 50 generations. If a chromosome that meets the fitness criteria is found early, it is returned; otherwise, the iteration continues until the final generation, after which the top chromosome based on fitness value is selected.

\subsection{Q-Learning (QL)}
Q-Learning is a model-free reinforcement learning algorithm that enables an agent to learn optimal actions within an environment by interacting with it. It operates by learning a Q-value function, which estimates the expected cumulative reward for taking a particular action in a given state and following an optimal policy thereafter. Through trial and error, Q-Learning updates these Q-values iteratively, using observed rewards to refine future decisions. This process allows the agent to eventually converge on an optimal policy, even without prior knowledge of the environment's dynamics.

QL in a given environment is characterized by two important variables: the state of the system/agent at any given time/episode and the action space of the system. While the state represents a snapshot of the environment at a given time, the action space of a QL algorithm is defined as the set of all possible actions an agent can take in a given environment, that change the agent's state and influence the rewards it receives, when executed. For our quantum control problem, the action space is represented by the set of possible amplitudes the pulse can have, where we limit this to be 100 values in the range [-1, 1]. 

The `state' of QL in our quantum control problem is represented by the state of the qubit at a given time in the evolution process. We use QL with epsilon-greedy strategy, where we explore the environment with a probability of $\epsilon$ and exploit the system by choosing the actions that have high Q-values  with a probability of 1-$\epsilon$. Because our quantum control problem has very short (i.e., few) time steps / episodes, the value of $\epsilon$ is set to 0.1 (10\% exploration and 90\% exploitation), to focus more on exploitation than exploration.  In this work, a learning rate of 0.001 and a reward decay factor of 0.9 are employed. The reward values are assigned as follows: -1 if the infidelity exceeds 0.5, 10 if the infidelity is below 0.5, 100 if the infidelity is below 0.1, and 500 if the infidelity is below 0.001. The initial actions are generated randomly. The maximum episodes the agent can take is limited to 500, and if the agent does not find an optimal path during this time, the state of the system at the end of the episode is returned.

\subsection{Deep Q-Network (DQN)}
Deep Q-Learning (DQN) is an extension of the Q-Learning algorithm that incorporates deep neural networks to handle environments with high-dimensional state spaces. While traditional Q-Learning relies on tabular representations (Q-table), DQN uses a neural network to approximate the Q-value function, enabling it to scale to more complex tasks. By integrating techniques like experience replay and target networks, DQN improves stability and convergence during training. 

The action space and state of the DQN algorithm are the same as that of the QL algorithm. The neural network we use is a simple 3 layer MLP (Multi Layer Perceptron) with hidden layers containing [64, 512, 256] units, respectively. Key hyper-parameters include a learning rate of 0.0001, an exploration fraction of 0.25, and a reward discount factor of 0.000001. Reward values are assigned as follows: 1 if the infidelity exceeds 0.5, 10 if the infidelity is below 0.5, 500 if the infidelity is below 0.1, and 5000 if the infidelity is below 0.001. The implementation details of our DQN algorithm are as follows:
\begin{itemize}
    \item We create a custom environment wrapped around gymnasium. Gymnasium~\cite{towers2024gymnasium} is a toolkit designed for developing single agent RL algorithms by providing environments for common RL experiments or the possibility to define custom environments, allowing researchers to easily design and run RL experiments tailored to specific needs.
    \item The neural net is implemented using Stable Baselines3. Stable Baseline3~\cite{stable-baselines3} is another popular library that implements state-of-the-art RL algorithms (like DQN and PPO) and simplifies the process of training, rapid experimentation and deploying RL models in Gymnasium environments.
    \item The input to the neural net is the state of the system at current time step and the output of the neural net is the optimal action for that state
\end{itemize}

\subsection{Proximal Policy Optimization (PPO)}
Proximal Policy Optimization (PPO) is a popular RL algorithm that strikes a balance between simplicity and efficiency. PPO simplifies the optimization process while maintaining stable updates by using a clipped objective function to prevent large, destabilizing policy updates, allowing for efficient learning across a variety of environments~\cite{schulman2017proximalpolicyoptimizationalgorithms}. PPO is widely used in RL due to its robustness, ease of implementation, and ability to perform well in both continuous and discrete action spaces.

Similar to DQN, we use Gymnasium to define a custom environment, where the action space, which consists of 100 discrete values within the range [-1, 1], is represented by the set of possible actions the qubit can take, and the `state' is represented by the state of the qubit at a given time. For the PPO algorithm, we use the implementation by Stable Baselines3, and used a 3 layer MLP with [64, 512, 256] units in each layer as the PPO's neural network. The following hyper-parameters are used in our experiments: a learning rate of 0.0001, entropy coefficient of 0.25, and a reward discount factor of 0.000001. Reward values are structured exactly the same to that of the DQN algorithm. For both DQN and PPO, initial actions are sampled from the action space randomly. Whereas DQN and PPO are highly effective algorithms for complex tasks, their use in simpler problems may be excessive. In such cases, the neural network could struggle to capture finer details, leading to sub-optimal learning.

\begin{figure}[htbp]
    \centering
    \begin{subfigure}[b]{0.29\textwidth}
        \centering
        \includegraphics[width=\textwidth]{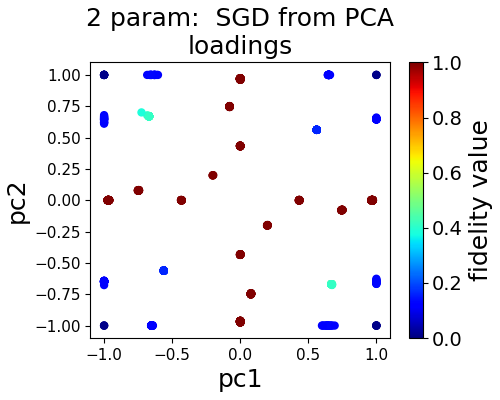} 
        \caption{}
        \label{fig:2paramPCASGD}
    \end{subfigure}
    \begin{subfigure}[b]{0.29\textwidth}
        \centering
        \includegraphics[width=\textwidth]{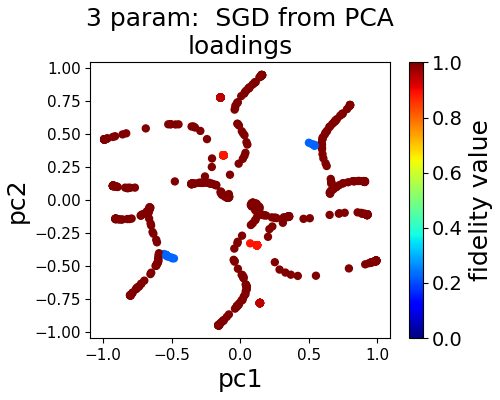} 
        \caption{}
        \label{fig:3paramPCASGD}
    \end{subfigure}
    \begin{subfigure}[b]{0.29\textwidth}
        \centering
        \includegraphics[width=\textwidth]{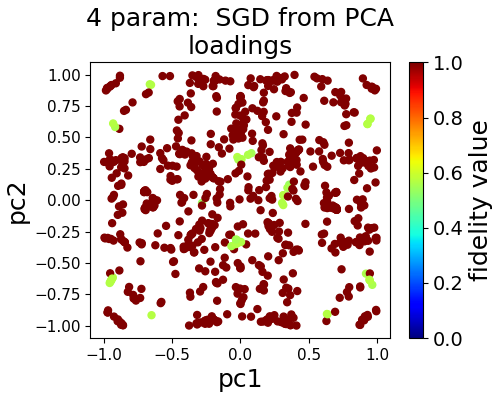} 
        \caption{}
        \label{fig:4paramPCASGD}
    \end{subfigure}
    
    \begin{subfigure}[b]{0.29\textwidth}
        \centering
        \includegraphics[width=\textwidth]{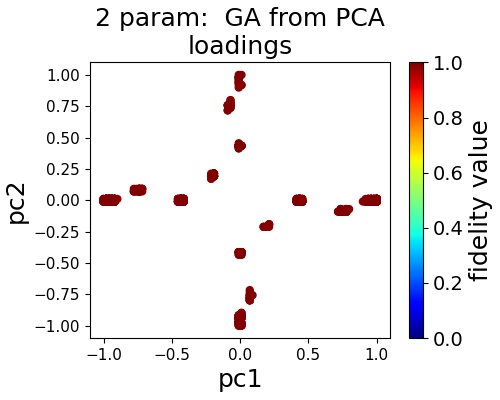} 
        \caption{}
        \label{fig:2paramPCAGA}
    \end{subfigure}
    \begin{subfigure}[b]{0.29\textwidth}
        \centering
        \includegraphics[width=\textwidth]{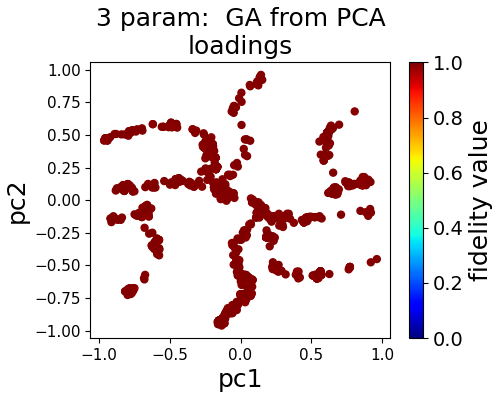} 
        \caption{}
        \label{fig:3paramPCAGA}
    \end{subfigure}
    \begin{subfigure}[b]{0.29\textwidth}
        \centering
        \includegraphics[width=\textwidth]{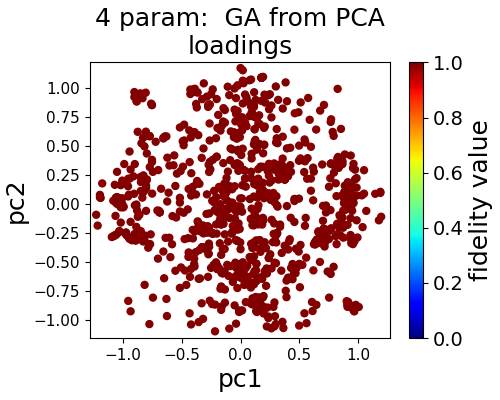} 
        \caption{}
        \label{fig:4paramPCAGA}
    \end{subfigure}
    \caption{The quantum landscape when using SGD and GA for 1000 tests and after applying PCA (a) SGD - 2 parameter, (b) SGD - 3 parameter, (c) SGD -  4 parameter, (d) GA - 2 parameter, (e) GA - 3 parameter, and (f) GA -  4 parameter}
    \label{fig:afterapplyingPCASGDGA}
\end{figure}


\section{Results Analysis and Discussion}
\label{sec:results}
In pursuit of discovering which algorithms among SGD, GA, QL, DQN and PPO are best at generating high fidelity control pulses and are able to explore the quantum landscape effectively, we conduct 1,000 tests for the 2, 3, and 4 parameter cases for each algorithm, and the resulting data points are passed through PCA to generate a 2D landscape.

Fig.~\ref{fig:afterapplyingPCASGDGA}(a-c) shows the results after running SGD for a maximum of 10,000 iterations. From the results in Fig.~\ref{fig:afterapplyingPCASGDGA}(a-c), we observe that even though most points in the two dimensional quantum landscape correspond to high fidelity values, there are a considerable number of points which are in the sub optimal (low fidelity) region even after giving the algorithm enough time to converge (10,000 iterations). The presence of sub-optimal points suggests that SGD can struggle with convergence to globally optimal solutions.

In Fig.~\ref{fig:afterapplyingPCASGDGA}(d-f), we present the resulting landscape when the approach used is GA. As can be seen in Fig.~\ref{fig:afterapplyingPCASGDGA}(d-f), and unlike SGD, the GA results display high fidelity pulses across all parameter cases. This indicates that GA has a stronger capability to explore the parameter space effectively and converge to optimal solutions. However, it is important to note that despite the favorable results, GA may still encounter issues if the initial population is poorly sampled. Random initialization can trap GA in sub-optimal regions, although such cases were not evident in our experiments. Overall, the GA demonstrates greater robustness and reliability compared to SGD in generating high-fidelity pulses.

\begin{figure}[htbp]
    \centering
    \begin{subfigure}[b]{0.29\textwidth}
        \centering
        \includegraphics[width=\textwidth]{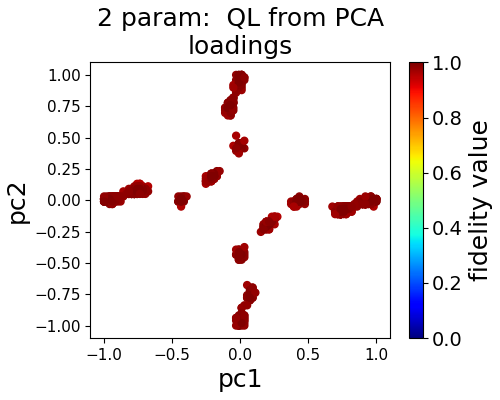} 
        \caption{}
        \label{fig:2paramPCAQL}
    \end{subfigure}
    \begin{subfigure}[b]{0.29\textwidth}
        \centering
        \includegraphics[width=\textwidth]{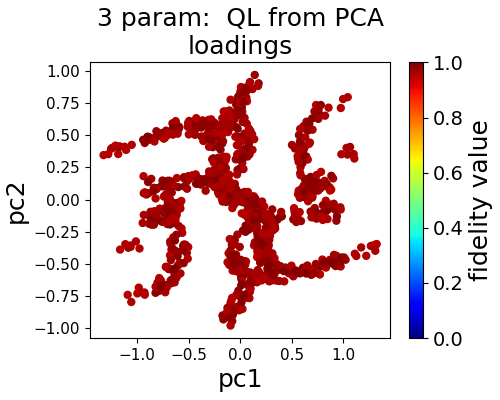} 
        \caption{}
        \label{fig:3paramPCAQL}
    \end{subfigure}
    \begin{subfigure}[b]{0.29\textwidth}
        \centering
        \includegraphics[width=\textwidth]{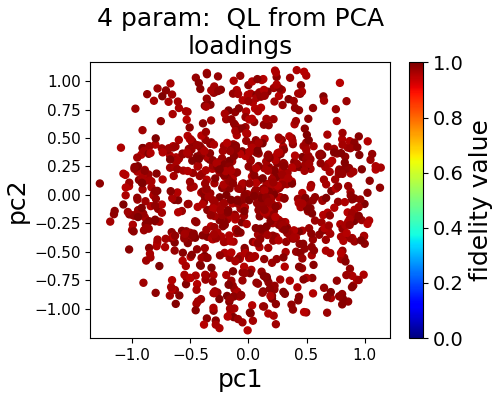} 
        \caption{}
        \label{fig:4paramPCAQL}
    \end{subfigure}
    \caption{The quantum landscape when using QL for 1000 tests and after applying PCA (a) 2 parameter, (b) 3 parameter, and (c) 4 parameter}
    \label{fig:afterapplyingPCAQL}
\end{figure}

The results of our experiment using QL, after passing through PCA, are presented in Fig.~\ref{fig:afterapplyingPCAQL}. When compared to SGD, QL produces a landscape that is close to the ideal brute-force landscape, with the data points concentrated in high-fidelity regions. This suggests that QL, similar to GA, is a more capable algorithm for navigating quantum control landscapes, as it consistently outperforms SGD in terms of generating optimal control pulses. This landscape demonstrates QL's potential for effective optimization in quantum control tasks.

In the experiments we conduct using DQN and PPO, the results reveal important insights into reward structuring of these algorithms. It was discovered that for problems requiring fewer steps per episode, as is the case for 2, 3, and 4 parameter quantum control, delayed rewards often fail to guide the model towards optimal solutions due to the shorter episode length. On the contrary,  the model tends to achieve better results when immediate rewards are prioritized, as the system has less time to accumulate delayed feedback effectively.
 
\begin{figure}[htbp]
    \centering
    \begin{subfigure}[b]{0.29\textwidth}
        \centering
        \includegraphics[width=\textwidth]{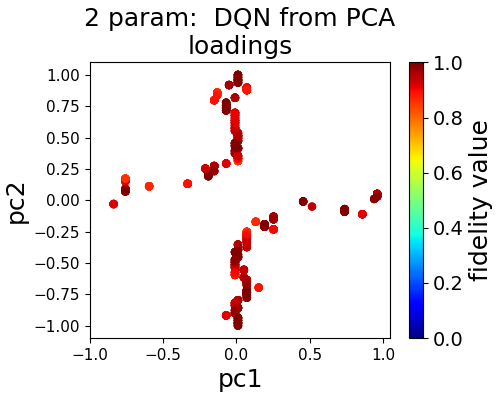} 
        \caption{}
        \label{fig:2paramPCADQN}
    \end{subfigure}
    \begin{subfigure}[b]{0.29\textwidth}
        \centering
        \includegraphics[width=\textwidth]{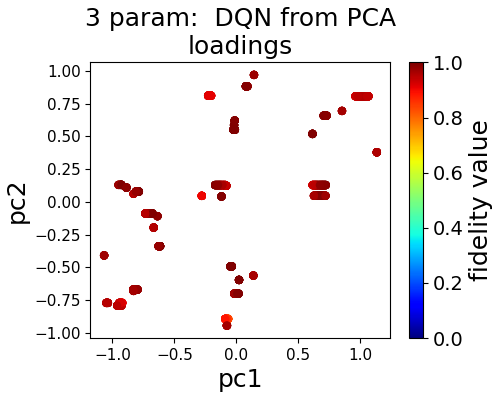} 
        \caption{}
        \label{fig:3paramPCADQN}
    \end{subfigure}
    \begin{subfigure}[b]{0.29\textwidth}
        \centering
        \includegraphics[width=\textwidth]{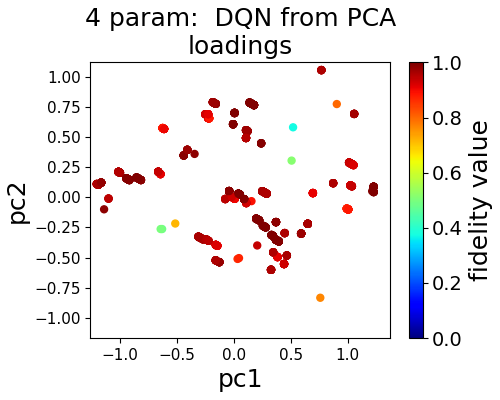} 
        \caption{}
        \label{fig:4paramPCADQN}
    \end{subfigure}

    \begin{subfigure}[b]{0.29\textwidth}
        \centering
        \includegraphics[width=\textwidth]{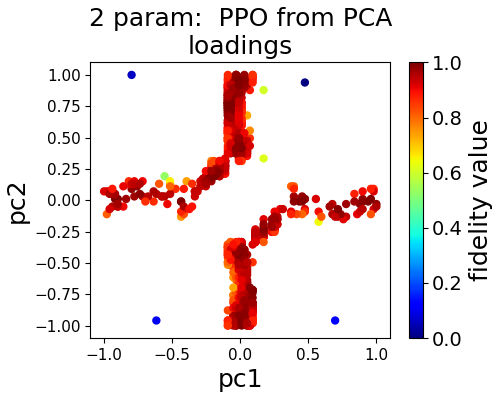} 
        \caption{}
        \label{fig:2paramPCAPPO}
    \end{subfigure}
    \begin{subfigure}[b]{0.29\textwidth}
        \centering
        \includegraphics[width=\textwidth]{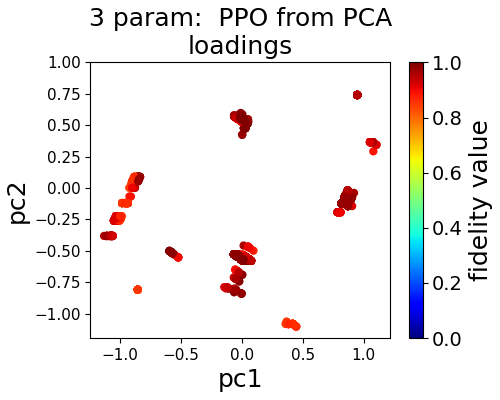} 
        \caption{}
        \label{fig:3paramPCAPPO}
    \end{subfigure}
    \begin{subfigure}[b]{0.29\textwidth}
        \centering
        \includegraphics[width=\textwidth]{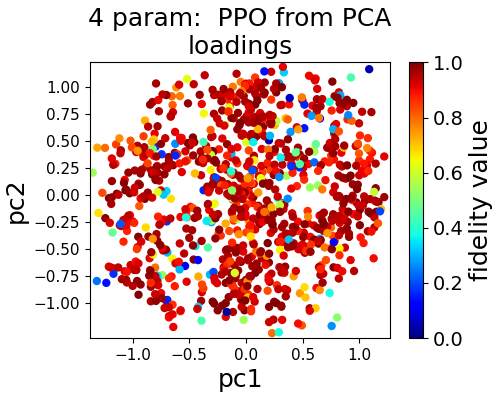} 
        \caption{}
        \label{fig:4paramPCAPPO}
    \end{subfigure}
    \caption{The quantum landscape when using DQN and PPO for 1000 tests and after applying PCA (a) 2 parameter-DQN, (b) 3 parameter-DQN, (c) 4 parameter-DQN (d) 2 parameter-PPO, (e) 3 parameter-PPO, and (f) 4 parameter-PPO}
    \label{fig:afterapplyingPCADQNPPO}
\end{figure}

As is the case with the other algorithms we discussed above, the experiments were repeated 1000 times and the results of our experiment using DQN and PPO, after passing through PCA,  are presented in Fig.~\ref{fig:afterapplyingPCADQNPPO}. From the results in Fig.~\ref{fig:afterapplyingPCADQNPPO}, we see that the DQN and PPO results align closely (with the exception of the 4 parameter case) with those of QL and GA, suggesting that reinforcement learning methods, when structured with appropriate reward schemes, offer strong potential for high-fidelity quantum control optimization. The few results in the sub-optimal region for both DQN and PPO (especially for the 4 parameter case), may suggest these algorithms might be an overkill if the problem is relatively simple as in our single-qubit state transition problem. One possible reason for this could be that the neural network in DQN and PPO fails to learn the relationships between its inputs and outputs as the quantum system being explored is fairly simple for a neural network - suggesting we might want to use advanced RL algorithms only when the traditional algorithms fail to generate high fidelity pulses. Overall, for our single-qubit state transition quantum control problem, GA and QL standout as the best performing algorithms.

\begin{figure}[h]
    \centering
    \begin{subfigure}[b]{0.3\textwidth}
        \centering
        \includegraphics[width=\textwidth]{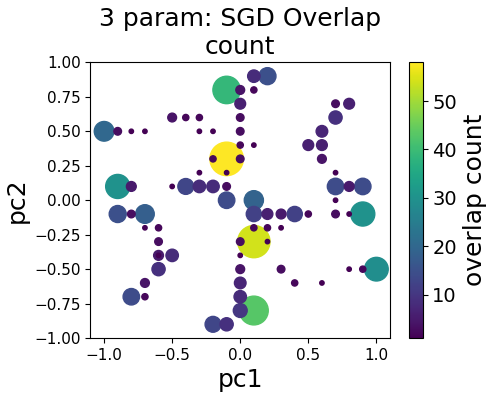} 
        \caption{}
        \label{fig:3paramSGDOVerlap}
    \end{subfigure}
    \begin{subfigure}[b]{0.3\textwidth}
        \centering
        \includegraphics[width=\textwidth]{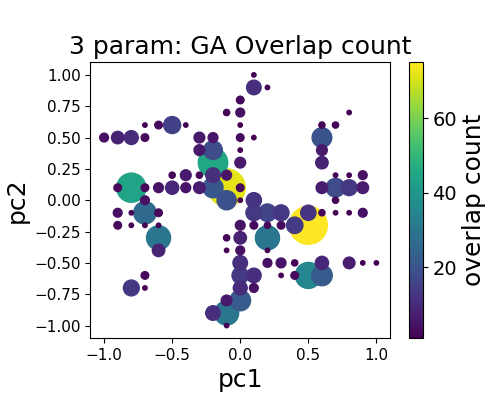} 
        \caption{}
        \label{fig:3paramGAOVerlap}
    \end{subfigure}
    \begin{subfigure}[b]{0.3\textwidth}
        \centering
        \includegraphics[width=\textwidth]{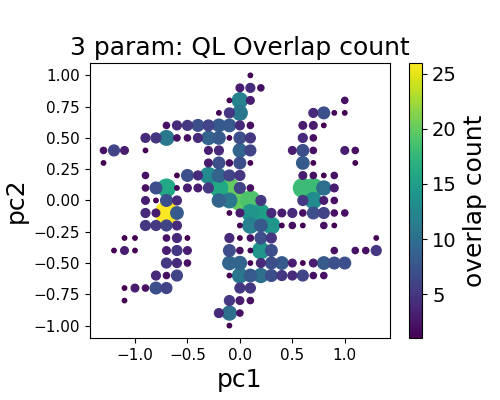} 
        \caption{}
        \label{fig:3paramQLOVerlap}
    \end{subfigure}
    \begin{subfigure}[b]{0.3\textwidth}
        \centering
        \includegraphics[width=\textwidth]{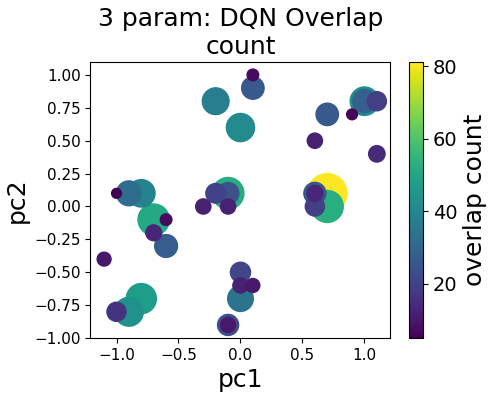} 
        \caption{}
        \label{fig:3paramDQNOVerlap}
    \end{subfigure}
    \begin{subfigure}[b]{0.3\textwidth}
        \centering
        \includegraphics[width=\textwidth]{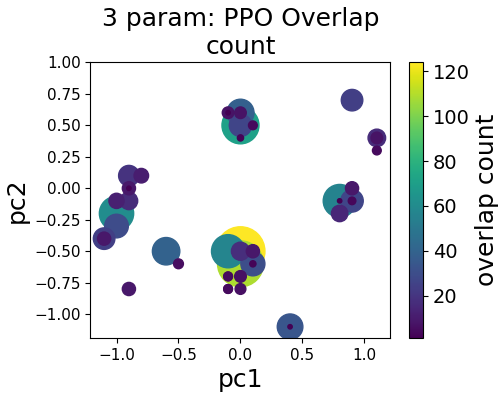} 
        \caption{}
        \label{fig:3paramPPOOVerlap}
    \end{subfigure}
    \caption{The overlap count for the 3 parameter case: the overlap count means the number of points in close proximity both in terms of coordinates and fidelity value. If certain number of points lie closer to each other in the 2D coordinate system and their fidelity values are close to each other, the overlap count in this region will be higher. Bigger circles represent higher count. (a) overlap count for 3 parameter-SGD, (b) overlap count for 3 parameter-GA, (c) overlap count for 3 parameter-QL (d) overlap count for 3 parameter-DQN, and (e) overlap count for 3 parameter-PPO}
    \label{fig:overlapSGDGAQLDQNPPO}
\end{figure}

Upon examining the landscape plots, it may initially seem that not all 1,000 test results are represented in each plot. This is due to overlapping points in the plots (i.e., where an approach finds the same solution multiple times), which creates the impression that fewer data points are present. For each algorithm and parameter count, we recorded the overlap of points to provide a comprehensive understanding. As an example, for the 3-parameter scenario, we present this overlap count in Fig.~\ref{fig:overlapSGDGAQLDQNPPO}. From this overlap plot, we see that while the points spread over the high fidelity regions for some algorithms, certain algorithms (DQN and PPO) tend to favor repeating specific combinations of inputs, resulting in clusters of points mainly in certain regions of the landscape. This clustering may suggest the tendency of certain algorithms to gravitate towards repeating certain pulses rather than exploring new regions of the parameter space.

Knowing the overlap count is relevant for a number of reasons: 
\begin{itemize}
    \item A high overlap count may indicate suboptimal results, suggesting that the algorithm may have missed high-fidelity regions, or
    \item it could imply that the quantum system is inherently complex, with limited solutions available to satisfy state transfer requirements, highlighting the need for more advanced control strategies.
\end{itemize}
On the other hand, if the overlap count remains low, it may imply that the algorithm has effectively explored and covered most high-fidelity regions in the solution space. This can indicate either that the algorithm is well-suited for the task or that the problem itself is relatively straightforward. In such cases, achieving successful state transfer may not require a complex or highly optimized quantum control strategy, as the solution can be reached with simpler techniques, saving computational resources and reducing algorithmic complexity.

\begin{figure}[H]
    \centering
    \begin{subfigure}[b]{0.33\textwidth}
        \centering
        \includegraphics[width=\textwidth]{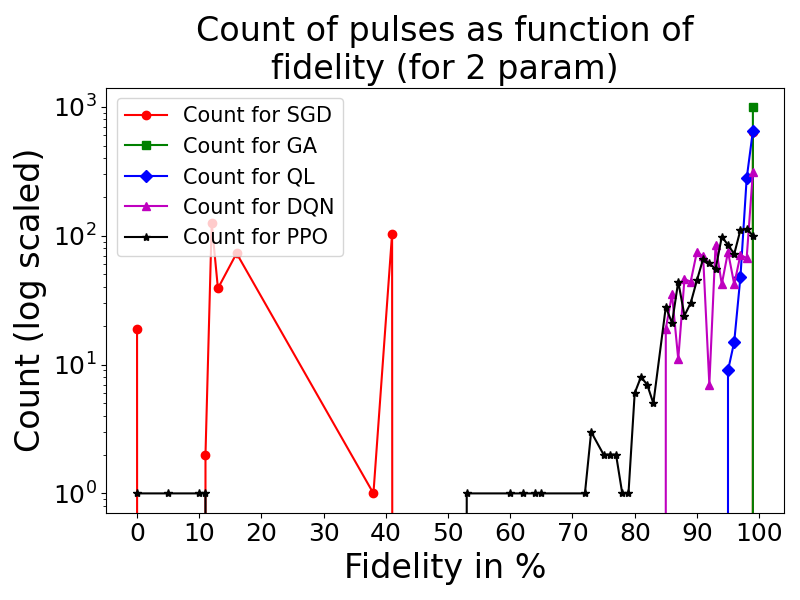} 
        \caption{}
        \label{fig:2paramOverlapAll}
    \end{subfigure}
    \begin{subfigure}[b]{0.33\textwidth}
        \centering
        \includegraphics[width=\textwidth]{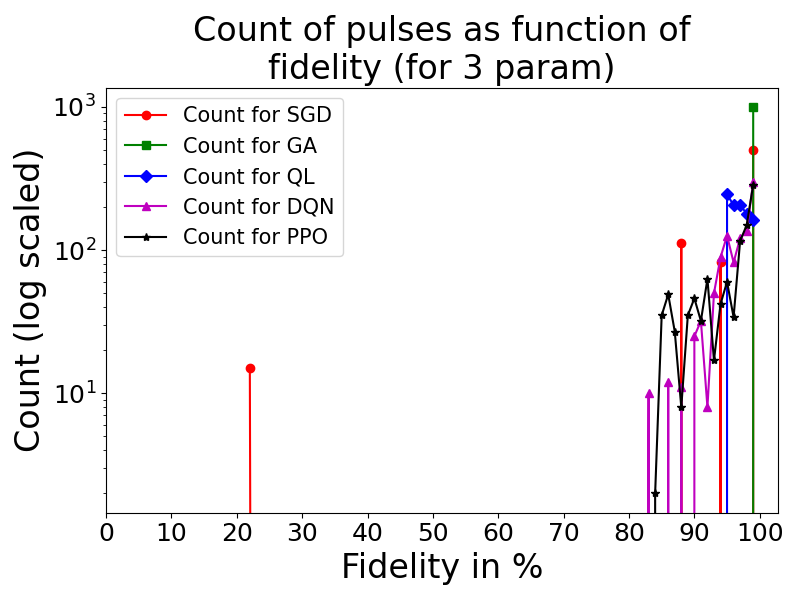} 
        \caption{}
        \label{fig:3paramOVerlapAll}
    \end{subfigure}
    \begin{subfigure}[b]{0.33\textwidth}
        \centering
        \includegraphics[width=\textwidth]{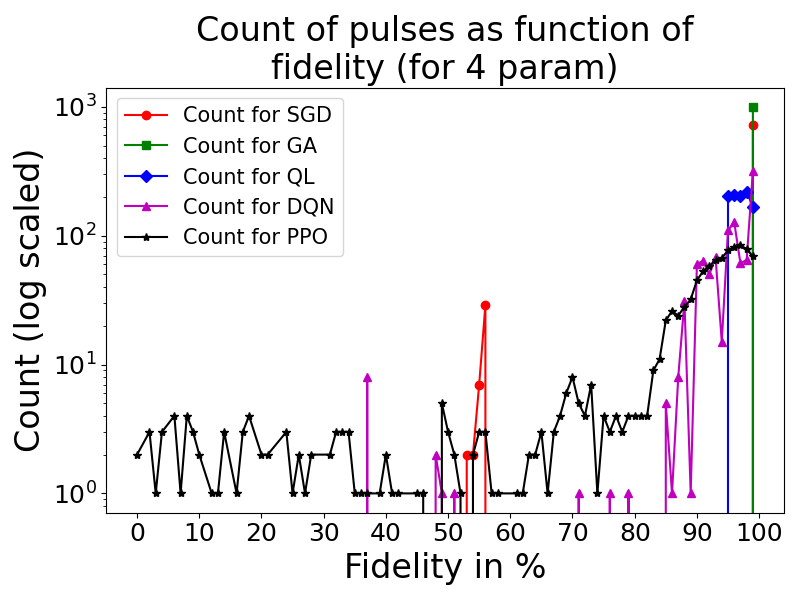} 
        \caption{}
        \label{fig:4paramOVerlapALL}
    \end{subfigure}
    
    \caption{Pulse counts for SGD, GA, QL , DQN and PPO algorithm as a function of fidelity: (a) 2 parameter, (b) 3 parameter, and (c) 4 parameter. Note that the Y axis values are given in log scale to show the distribution of points more clearly }
    \label{fig:overlapPlotAll}
\end{figure}

To quantitatively evaluate how effectively each algorithm identifies pulses that result in high fidelity, we present the distribution of pulse counts as a function of fidelity in Fig.~\ref{fig:overlapPlotAll}. From the plots in Fig.~\ref{fig:2paramOverlapAll} we observe that 2-parameter SGD algorithm has many points concentrated in the low-fidelity region, indicating it struggles to find high-fidelity solutions. In contrast, GA has all its pulses within the high-fidelity region (close to 100\%), indicating it is highly effective in achieving high fidelity with two parameters. QL, DQN, and PPO display a mixed distribution, with most of their points also leaning toward higher fidelity, but they are less consistent than GA. For the 3-parameter case in Fig.~\ref{fig:3paramOVerlapAll}, SGD continues to struggle, GA maintains its dominance, achieving high fidelity with all pulses. QL, DQN, and PPO are distributed across high-fidelity regions as well, with DQN and PPO clustering closer to higher fidelity. In Fig.~\ref{fig:4paramOVerlapALL}, which explores the 4-parameter scenario, a more spread-out distribution is observed. SGD still has many points scattered across lower fidelity values, showing it is less efficient even with additional parameters. GA remains highly effective, with all points concentrated near 100\% fidelity. Interestingly, QL, DQN, and PPO also show a wider range in fidelity, but most points are in the high-fidelity region. PPO, in particular, has few pulses distributed across a wider fidelity range compared to GA. 
In summary, an algorithm capable of generating a wide variety of high fidelity pulses is preferred over one that produces only a limited set, as having a broader selection of high fidelity pulses provides flexibility. This range is valuable since some pulses may not be easily realizable, making it essential to have alternative pulses that enhance the robustness and applicability of a control algorithm.

\section{Solution  Space Complexity}
\label{sec:solutionspacecomplexity}

In the previous section, we have seen that some approaches have the potential to yield more (i.e., increased variety) high-fidelity solutions, and others can fall into local optima traps. Yet, the question remains, how should we evaluate the complexity of a QCL in order to make informed choices regarding which approach(es) to use? In order to study the complexity of the solution space, we analyse how sparse/dense the landscapes are using clustering algorithms and distance measures. To find clusters in the solution space, we employ a cluster number agnostic algorithm-DBSCAN~\cite{ester1996density}, and to find the distances between points within a cluster;  we use the Euclidean distance. If two clusters are of comparable area and the distance between the points within one cluster is smaller than the distance between points within the other cluster, this tells us the first cluster is more dense. This, in turn, can reveal that an approach is more/less likely to find a high-fidelity solution in this region of the landscape. 

To measure how dense the landscape is, we define a term we called cluster density index (CDI), which is the inverse of how sparse a region is,  as:
\begin{equation}
\text{CDI} = \frac{\bar{A}}{\bar{D}}
\label{equ:sparsity}
\end{equation}
where, \textit{CDI} refers to the average Cluster Density Index, $\bar{A}$ is the average area of clusters calculated as mean of individual cluster areas which are calculated using Delaunay Triangulation~\cite{10.1145/235815.235821}, and $\bar{D}$ is the average, over all clusters, of the mean pairwise distances between points in a given cluster. In Table~\ref{tab:sparsityIndex}, $\bar{L}$ refers to the average inter-cluster distance.
\begin{table*} [htbp]
    \centering
    \caption{Cluster Density Index: describes how sparse / dense a landscape is}
    \begin{adjustbox}{max width=\textwidth}
    \begin{tabular}{|>{\centering\arraybackslash}m{0.041\linewidth}|>{\centering\arraybackslash}m{0.041\linewidth}|>{\centering\arraybackslash}m{0.041\linewidth}|>{\centering\arraybackslash}m{0.041\linewidth}|>{\centering\arraybackslash}m{0.041\linewidth}|>{\centering\arraybackslash}m{0.041\linewidth}|>{\centering\arraybackslash}m{0.041\linewidth}|>{\centering\arraybackslash}m{0.041\linewidth}|>{\centering\arraybackslash}m{0.041\linewidth}|>{\centering\arraybackslash}m{0.041\linewidth}|>{\centering\arraybackslash}m{0.041\linewidth}|>{\centering\arraybackslash}m{0.041\linewidth}|>{\centering\arraybackslash}m{0.041\linewidth}|>{\centering\arraybackslash}m{0.041\linewidth}|>{\centering\arraybackslash}m{0.041\linewidth}|>{\centering\arraybackslash}m{0.041\linewidth}|}  
    \hline
          \rule{0pt}{4.5ex} \multirow{2}{*}{} &  \multicolumn{3}{c|}{\textbf{SGD}} & \multicolumn{3}{c|}{\textbf{GA}} & \multicolumn{3}{c|}{\textbf{QL}} & \multicolumn{3}{c|}{\textbf{DQN}} & \multicolumn{3}{c|}{\textbf{PPO}} \\  \cline{2-16} 
          \rule{0pt}{5.5ex}& 2 param & 3 param & 4 param & 2 param & 3 param & 4 param & 2 param & 3 param & 4 param & 2 param & 3 param & 4 param & 2 param & 3 param & 4 param\\ \hline
         $\bar{L}$ & \gradient{0.9463}& \gradient{1.0501}& \gradient{1.2232}& \gradient{0.9391}& \gradient{0.9576}& \gradient{1.2380}& \gradient{0.8398}& \gradient{1.4588}& \gradient{1.3698}& \gradient{0.8919}& \gradient{1.1551}& \gradient{0.9895}& \gradient{0.8836}& \gradient{1.2656}& \gradient{1.1495}\\ \hline
         $\bar{D}$ & \gradient{0.0035}& \gradient{0.1280}& \gradient{0.1392}& \gradient{0.0234}& \gradient{0.1094}& \gradient{0.1418}& \gradient{0.0697}& \gradient{0.2178}& \gradient{0.1966}& \gradient{0.0423}& \gradient{0.0167}& \gradient{0.0253}& \gradient{0.1383}& \gradient{0.0273}& \gradient{0.1593}\\ \hline
         $\bar{A}$ & \gradient{0.0001}& \gradient{0.0251}& \gradient{0.1023}& \gradient{0.0010}& \gradient{0.0505}& \gradient{0.2282}& \gradient{0.0133}& \gradient{0.4129}& \gradient{0.4605}& \gradient{0.0036}& \gradient{0.0024}& \gradient{0.0027}& \gradient{0.0333}& \gradient{0.0185}& \gradient{0.0724}\\ \hline
         CDI & \gradient{0.0075}& \gradient{0.1962}& \gradient{0.7346}& \gradient{0.0449}& \gradient{0.4617}& \gradient{1.6096}& \gradient{0.1902}& \gradient{1.8957}& \gradient{2.3420}& \gradient{0.0844}& \gradient{0.1427}& \gradient{0.1075}& \gradient{0.2404}& \gradient{0.6766}& \gradient{0.4543}\\ \hline
    \end{tabular}
    \end{adjustbox}
    \label{tab:sparsityIndex}
\end{table*}

Using~\eqref{equ:sparsity}, we calculate and present the resulting Cluster Density Index in Table~\ref{tab:sparsityIndex}. A high CDI value is desirable, because a high CDI (according to~\eqref{equ:sparsity}), ideally, means there are high fidelity regions in the landscape which have large and dense clusters (it is worth to note that the landscapes are generated after using PCA on the output of the different algorithms). To increase the CDI value, $\bar{A}$ has to be high while $\bar{D}$ is kept low. Typically, higher CDI means there are many high fidelity points (higher cluster area) in the landscape with the distance between individual points being low (individual points are closer to each other in the cluster). The results in Table~\ref{tab:sparsityIndex} confirm that, in general, algorithms with larger clusters in the landscape, have higher cluster density index values. The best performing algorithms (GA and QL) have the highest CDI as compared to the remaining algorithms. When we compare the results in terms of parameter count, for the same algorithm, we see that for SGD, GA and QL increasing the number of parameters results in an increased CDI value. For DQN and PPO, 3 parameter pulses seem to be better than both 2 parameter and 4 parameter pulses - a detailed future study could be necessary to further investigate this outlier behavior.

However, it should be noted that a higher value of CDI does not necessarily mean that there are many high fidelity points in a given region - because we can get high CDI value even when $\bar{A}$ is very small (extremely few points in each cluster) provided that $\bar{D}$ is even smaller. For example, looking at the CDI value in Table~\ref{tab:sparsityIndex}, for 3 parameter PPO, it is relatively larger than the value for 2 parameter PPO. This is because the average cluster area is very small for 3 parameter PPO, but the average distance between the clusters is even smaller, resulting in a relatively higher CDI value for the 3 parameter PPO than that of 2 parameter. However, this does not mean 3 parameter PPO outperforms 2 parameter PPO in terms of exploring the QCL and generating control pulses with high fidelity (indeed a quick check of Fig.\ref{fig:afterapplyingPCADQNPPO} (d) and (e),  can reveal that 2 parameter PPO results are better at exploring the landscape when compared to 3 parameter case). While higher CDI values can sometimes result from small cluster areas paired with even smaller inter-point distances, these instances represent edge cases. In most scenarios, CDI serves as a reliable metric for assessing cluster quality, as it typically indicates regions of high fidelity with tightly packed points (we refer, for example, CDI results in Table~\ref{tab:sparsityIndex} for GA and QL and verify results by looking at Fig.~\ref{fig:afterapplyingPCASGDGA}(d-f) and Fig.~\ref{fig:afterapplyingPCAQL}). In general, CDI captures the relation between cluster area and point distribution, making it a valuable measure for solution space complexity analysis.

The CDI can be a very powerful tool for studying the QCL even at higher dimensions than we have considered. Because DBSCAN can work with higher dimensional data, and Euclidean distances can be calculated for any dimensions, CDI can potentially allow us to understand how probable it is to find high fidelity clusters in a landscape without even using dimensionality reduction techniques. 

\section{Conclusion}
In this work, we studied the dynamics of quantum control landscapes through the use of PCA as a dimensionality reduction technique, to visualize higher-dimensional quantum landscapes and algorithms best suited for solving quantum control problem. Our findings reveal that dimensionality reduction techniques are important to analyze the intricate nature of high dimensional quantum control. In addition, we noted that increasing the number of parameters (dividing the total evolution time into more discrete parts) yields landscapes with more high-fidelity regions. However, this does not guarantee a reduction in low-fidelity areas; rather, it enhances the chances of landing on high-fidelity regions when the number of parameters is increased. It should be noted that all dimensionality reduction techniques involve some degree of information loss; as such it would be highly relevant to compare and contrast characteristics of other dimensionality reduction techniques when applied to QCL.

Comparative analyses of traditional and machine learning algorithms in solving the quantum control problem for the given Hamiltonian demonstrates that while Stochastic Gradient Descent (SGD) can perform poorly, Genetic Algorithms (GA) excel at finding high fidelity pulses and exploring the landscape well. Among machine learning approaches, Q-learning (QL) showed promising results, while Deep Q-Network (DQN) and Proximal Policy Optimization (PPO) were less effective, suggesting that simpler algorithms may be advantageous for straightforward problems. Additionally, the design of the reward function in DQN and PPO significantly impacts performance, with immediate rewards being more beneficial for short episode tasks.

The solution space complexity analysis through the use of cluster density index, indicates that both GA and QL achieved high cluster-density scores for 4-parameter quantum control - meaning those algorithms result in a larger QCL clusters with closely spaced individual points. When comparing the results in terms of parameter count, the cluster-density-score increase when the parameter count is increased from 2 parameter to 3 parameter and then to 4 parameter  with exceptions noted for DQN and PPO. 

Our results demonstrate that dimensionality reduction tools, in particular PCA, can be highly effective in capturing the relevant features of the control landscape. While we have restricted our analysis to a small number of parameters to ensure that we can easily access the full QCL via brute force for comparison, it is worth highlighting that many of the considered algorithms can accommodate optimizing a much larger number of parameters~\cite{CoopmansPRR} or be applied to more complex multi-qubit systems. Our framework for characterising the solution space can nevertheless be readily applied without incurring significant additional computational overhead. The obvious caveat then being that achieving control over these more complex systems will necessitate high dimensional control pulses and, therefore, it still remains to determine the tradeoff between how many principal components are required to maintain the important features of the QCL. Furthermore, our results help to get a deeper understanding of quantum control strategies, emphasizing the importance of parameter selection and reward design in optimizing algorithm performance.

Future research could explore why certain algorithms (example: DQN and PPO) under-perform in short RL episodes, and identify the types of problems that require RL versus those better suited for traditional algorithms. Further research could also establish clear metrics to distinguish between simple and complex problems, helping determine the most appropriate algorithms based on problem complexity and required computational resources.

\section*{Data availability}
The data generated during this study are available from the corresponding author upon reasonable request.

\section*{Code availability}
The code used in this work is accessible on GitHub through the following link: 
\href{https://github.com/Hafdream/Quantum-control-and-landscape}{click here for code}

\section*{Acknowledgements}

This publication has emanated from research conducted with the financial support of Science Foundation Ireland under Grant number 18/CRT/6183. For the purpose of Open Access, the author has applied a CC BY public copyright license to any Author Accepted Manuscript version arising from this submission.




\end{document}